\documentclass[twocolumn,showpacs,preprintnumbers,amsmath,amssymb]{revtex4-1}
\usepackage{graphicx}
\usepackage{textcomp}

\begin{document}
\title{Topological phase transition in non-Hermitian quasicrystals}
  \normalsize

\author{S. Longhi$^{1,2\ast}$}
\address{$^1$ Dipartimento di Fisica - Politecnico di Milano, Piazza Leonardo da Vinci 32, 20133 Milan, Italy}
\address{$^2$ Istituto di Fotonica e Nanotecnologie - Consiglio Nazionale delle Ricerche, Piazza Leonardo da Vinci 32, 20133 Milan, Italy}

%
\bigskip
\begin{abstract}
The discovery of topological phases in non-Hermitian open classical and quantum systems challenges our current understanding of topological order.  Non-Hermitian systems exhibit unique features with no counterparts in topological Hermitian models, such as failure of the conventional bulk-boundary correspondence and non-Hermitian skin effect. Advances in the understanding of the topological properties of non-Hermitian lattices with translational invariance have been reported in several recent studies, however little is known about non-Hermitian quasicrystals. Here we disclose topological phases in a quasicrystal with parity-time ($\mathcal{PT}$) symmetry, described by a non-Hermitian extension of the Aubry-Andr\'e-Harper model. It is shown that the metal-insulating phase transition, observed at the  $\mathcal{PT}$ symmetry breaking point, is of topological nature and can be expressed in terms of a winding number. A photonic realization of a non-Hermitian quasicrystal is also suggested.
\end{abstract}

\maketitle
{\it Introduction.} The discovery of topological phases of matter has introduced a major twist in condensed matter physics \cite{R1,R2} with great impact in other areas of physics, such as photonics, atom optics, acoustics and mechanics  \cite{R3,R4,R5,R6,R6bis,R6tris}. Topological band theory  classifies Hermitian topological systems depending on their dimensionality and symmetries \cite{R7,R8}. The bulk topological invariants are uniquely reflected in robust edge states localized at open boundaries. The ability to engineer non-Hermitian Hamiltonians, demonstrated in a series of recent experiments \cite{R9,R10,R11,R12,R13,R14,R15}, and the related observation of unconventional topological boundary modes sparked a great interest to extend topological band theory to open systems
\cite{R16,R17,R17bis,R18,R19,R20,R21,R24,R25,R27,R29,R30,R31,R32,R33,R34,R36,R37,R38,R39,R39bis,R39tris,R40}. Striking features are the  failure of the conventional bulk-boundary correspondence \cite{R27,R32,R37,R39,R39bis,R39tris}, eigenstate condensation \cite{R31,R32}, non-Hermitian {\em skin effect} \cite{R34,R39,R39bis}, and the sensitivity of the bulk spectra on boundary conditions  \cite{R17,R38,R39tris,R40,R41}.\\ 
Most of previous studies have concerned with crystals, however little is know about topological properties of non-Hermitian quasicrystals. Quasicrystals (QCs) constitute an intermediate
phase between fully periodic lattices and fully disordered media, showing a long-range order but no periodicity \cite{R42,R43}. A paradigmatic model of a one-dimensional (1D) QC is provided by the Aubry-Andr\'e-Harper (AAH) Hamiltonian \cite{R43,R44,R45}, which  is known to show a metal-insulator phase transition \cite{R44,R45,R46}. In the Hermitian case, the  AAH Hamiltonian is topologically nontrivial because it can be mapped into a two-dimensional quantum Hall system on a square lattice \cite{R47,R48,R49,R50}. A few recent studies have considered some non-Hermitian extensions of  the  AAH model \cite{R22bis,R22,R23,R26,R28,R35}, mainly with a commensurate potential and with open boundary conditions. Such numerical studies investigated how gain and loss distributions affect edge states and parity-time ($\mathcal{PT}$) symmetry breaking \cite{R22bis,R22,R23,R28}, the Hofstadter butterfly spectrum \cite{R22},  and the localization properties of eigenstates \cite{R26,R35}. However, so far there is not any evidence of {\em{topological}} phases and {\it topological} phase transitions in non-Hermitian QCs.\par
The aim of this Letter it to disclose topological phases in non-Hermitian quasicrystals, described by a  $\mathcal{PT}$-symmetric extension of the AAH model. The main result is that  the localization-delocalization phase transition, observed at the  $\mathcal{PT}$ symmetry breaking point in the thermodynamic limit,  is of topological nature and can be expressed in terms of a winding number which characterizes the two distinct phases of the system.  A photonic realization of the topological phase transition is proposed, which is based on the spectral properties of mode-locked lasers with an intracavity etalon. \par
{\it Non-Hermitian Aubry-Andr\'e-Harper model.} The tight-binding Hamiltonian of the AAH model, describing the hopping dynamics on a 1D lattice with an incommensurate potential, reads \cite{R42,R43,R44,R45,R46}
\begin{equation}
H(\varphi) \psi_n=J(\psi_{n+1}+\psi_{n-1})+V_n \psi_n 
\end{equation}
for the occupation amplitudes $\psi_n$ at the various sites of the lattice, where $J$ is the hopping rate,  
\begin{equation}
V_n=V \cos( 2 \pi \alpha n + \varphi)
\end{equation}
is the onsite potential, $V$ and $\varphi$ are the amplitude and phase of the potential, and $\alpha$ is irrational for a QC.  Here we consider a non-Hermitian extension of the AAH Hamiltonian by complexification of the potential phase $\varphi$, i.e we assume
\begin{equation}
\varphi=\theta+i h
\end{equation}
which yields
\begin{equation}
V_n  =  V \cos(2 \pi \alpha n+\theta+i h) 
\end{equation}
for the incommensurate on-site potential. Note that for $\theta=0$ one has $V_{-n}=-V_n^*$ and $H(\varphi)$ is $\mathcal{PT}$ symmetric.  In the Hermitian limit $h=0$, $H$ is an almost Mathieu operator which shows very rich and subtle spectral features, which have been extensively investigated in the mathematical literature \cite{Martini,Martini1}. Roughly speaking, for irrational $\alpha$ the Hamiltonian $H$ shows a metal-insulator phase transition: for $V<2J$ all eigenstates are delocalized (metallic phase) and the spectrum is independent of $\theta$, whereas for $V>2J$ all eigenstates are localized (insulating phase) with an inverse localization length (Lyapunov exponent) independent of the eigenenergy and given by $\gamma={\rm log}(V/ 2 J)$ \cite{R44}. Here we are interested to study the bulk properties of $H$ with complex phase $\varphi$ and to disclose topological phases in the thermodynamic limit $L \rightarrow \infty$ of number $L$ of lattice sites. Since any irrational number $\alpha$ can be approximated by a sequence of rational numbers $p_n/q_n$ with $p_n,q_n$ prime integers and $p_n,q_n \rightarrow \infty$ as $n \rightarrow \infty$ \cite{cazz}, in numerical simulations one can assume as usual a finite  (yet arbitrarily large) number of sites $L=q_n$ on a ring with periodic boundary conditions $\psi_{n+L}=\psi_n$. Clearly, in an ordinary crystal ($\alpha$ rational) in the thermodynamic limit $L \rightarrow \infty$ a large number of unit cells are reproduced inside the ring, however for irrational $\alpha$ there is not any unit cell that is exactly reproduced inside the ring.\par
{\it Topological phases, symmetry breaking and metal-insulator transition.}  
Topological properties of 1D superlattices and QCs in Hermitian models have attracted great interest recently \cite{R47,R48,R50,R51,R52,R53,R53bis,R53uff,R53tris,bastaa}. In particular, the connection between the Hermitian AAH model and the two-dimensional quantum Hall (Harper-Hofstadter) system \cite{R47,R53uff}, when the phase $\theta$ is considered as a synthetic additional dimension, has lead to the demonstration of topological pumping \cite{Tho} of edge states in photonic QCs \cite{R47}. However, the possibility that topological properties of 1D QCs can emerge from higher dimensions is a matter of debate \cite{R46,R51,R52,R53uff}.
In a recent work \cite{R39}, Gong and coworkers introduced a topological classification of non-Hermitian Hamiltonians that can be safely applied to systems with broken translational invariance. Following such an approach, we consider here the case $V<2 J$, corresponding to the delocalized phase in the Hermitian limit $h=0$, and assume the imaginary phase term $h$ as the control (deformation) parameter of the Hamiltonian $H(\varphi)=H(\theta,h)$, where the dependence of $H$ on $\theta$ and $h$ is defined by Eqs.(1-3).  For $h=0$, the spectrum of $H(\theta)$ is absolutely continuous, has a Cantor-set structure with a dense set of gaps and is independent of $\theta$ \cite{Martini,Martini1}. For example, when $\alpha$ is the inverse of the golden mean, $\alpha=(\sqrt{5}-1)/2$, the spectrum consists of mainly three $^\prime$bands$^\prime$,
which again consist of three sub-bands and so forth [see upper panel in Fig.1(a)]. Let $E_B$ be a base energy which is not an eigenenergy of $H$ but it is assumed to be embedded in a small gap of the Cantor set. Let us then introduce a winding number $w=w(h)$ as follows
\begin{equation}
w(h)=\lim_{L \rightarrow \infty} \frac{1}{2 \pi i} \int_0^{2 \pi} d \theta \frac{\partial}{\partial \theta} {\rm log} \;  {\rm det} \left\{ H \left( \frac{\theta}{L},h \right)-E_B  \right\}
\end{equation} 
The winding number $w$ counts the number of times the complex spectral trajectory encircles the base point  $E_B$ when the real phase $\theta$ varies from zero to $ 2 \pi$ \cite{R39}.
 As shown in the Supplemental Material \cite{supp}, $w=w(h)$ does not depend on $E_B$, is quantized and can take only two values depending on the strength of the non-Hermitian phase $h$
 \begin{equation}
 w= \left\{
 \begin{array}{cc}
 0 & h<h_c \\
 -1 & h>h_c
 \end{array}
 \right.
 \end{equation}
where the critical value $h_c$ is given by
\begin{equation}
h_c= \log \left( \frac{2J}{V} \right).
\end{equation}
This shows that there are two distinct topological phases of $H$. The two phases correspond to entirely delocalized eigenstates and real energy spectrum (unbroken $\mathcal{PT}$ phase) for $w=0$, and to entirely localized eigenstates and complex energy spectrum (broken $\mathcal{PT}$ phase) for $w=-1$. Such a result thus shows that the localization and symmetry-breaking phase transitions at $h=h_c$ are of topological nature.\\ 
\begin{figure*}
\includegraphics[width=17cm]{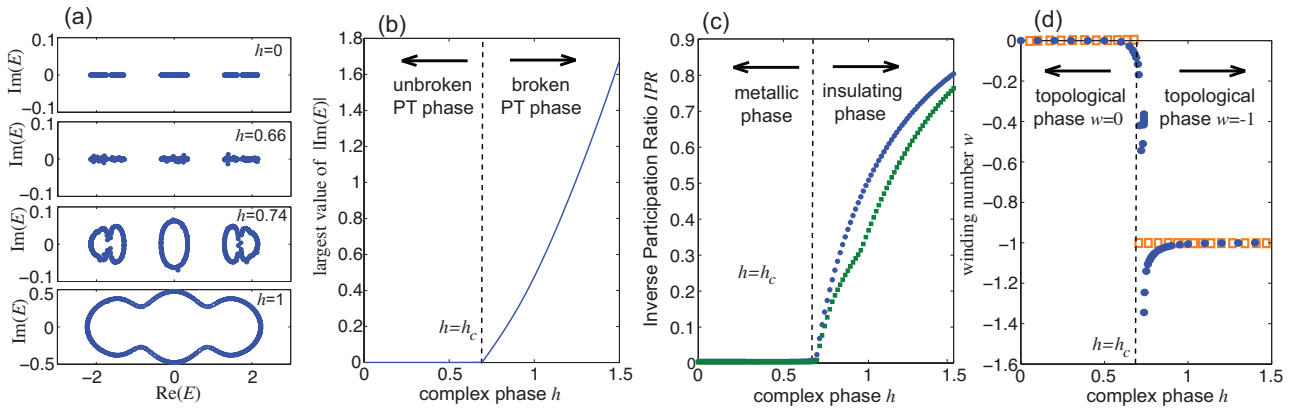}
\caption{(Color online) Topological phase transition in the non-Hermitian Aubry-Andr\'e-Harper  Hamiltonian $H(\theta,h)$ for $J=V=1$, $\alpha=(\sqrt{5}-1)/2$. Periodic boundary conditions are assumed on a lattice with $L=610$ sites. (a) Energy spectrum $E$ of $H$ for $\theta=0$ and for a few increasing values of the complex phase $h$. (b) Behavior of the largest value of $|{\rm Im}(E)|$ versus $h$ for $\theta=0$. The dashed vertical curve corresponds to the critical value $h_c={\rm log}(2J/V) \simeq 0.6931$ predicted in the thermodynamic limit $L \rightarrow \infty$. (c) Numerically-computed behavior of the largest and smallest values of inverse participation ratio IPR of eigenstates versus $h$. (d) Behavior of the winding number $w$ versus $h$ for $E_B$ as computed using Eq.(5) (circles) and Eq.(S-12) given in the Supplemental material \cite{supp} (squares). The base energy $E_B$ used to compute the winding number $w$ is $E_B=0$. The value of $w$ does not change if the base energy $E_B$ is varied, provided that it remains embedded in the central $^{\prime}$band$^{\prime}$  of the Cantor set of $H$ at $h=0$.}
\end{figure*}
The detailed proof of the above statements is given in the Supplemental Material \cite{supp}. Here we just briefly outline the main two steps of the proof. As a first step, by a similarity transformation the Hamiltonian $H(\theta,h)$ is connected to the Hatano-Nelson model \cite{R16,R17,R18} with incommensurate potential, and the localization properties of the eigenstates of $H$ for $h \neq 0$ are thus determined from those of the AAH model in the Hermitian limit $h=0$ (Sec.S.1 in \cite{supp}). This also entails that the non-Hermitian delocalization/localization phase transition at $h=h_c$ coincides with the $\mathcal{PT}$ symmetry breaking phase transition in the thermodynamic limit (the energy spectrum is entirely real in the delocalized phase, while it becomes complex in the localized phase). As a second step, the topological nature of the phase transition is demonstrated by direct computation of the winding number $w$ (Sec.S.-2 in \cite{supp}), which involves some mathematical steps and a result derived by Thouless \cite{R54} that relates the localization length of eigenstates and the density of states.\\
To exemplify and check the validity of the analytical results, we performed numerical diagonalization analysis of the matrix Hamiltonian $H$ by varying the non-Hermitian phase $h$ assuming $\alpha=(\sqrt{5}-1)/2$, $L=610$ (corresponding to $\alpha L \simeq 377$), $J=V=1$, and periodic boundary conditions. The localization of eigenstates is measured by the inverse of the participation ratio $IPR=\sum_{n}|\psi_n|^4 / (\sum_n |\psi_n|^2)^2$, with $IPR \simeq 1/L \simeq 0$ for a delocalized state and $IPR \simeq 1$ for a fully localized state. Figure 1(a) shows a few examples of the energy spectrum $E$ of $H$ for a few increasing values of the non-Hermitian phase $h$, whereas Fig.1(b) shows the behavior of the largest value of $|{\rm Im}(E)|$ versus $h$. A rather abrupt increase of ${\rm max}\{ |{\rm Im}(E) |\}$ from zero is clearly observed near $h=h_c \simeq 0.6931$, corresponding to the critical value of the complex phase predicted by the theoretical analysis [Eq.(7)]. The behavior of the IPR versus $h$, for the eigenstates with either the largest or smallest IPR, is depicted in Fig.1(c). Clearly, according to the theoretical analysis the critical value $h=h_c$ separates the metal and insulating phases. The topological nature of the phase transition is illustrated in Fig.1(d), where the winding number $w$ versus $h$ is numerically computed using Eq.(5). Note that contrary to the rather general result of Ref.\cite{R39}, the trivial topological phase  $w=0$ corresponds here to all eigenstates being delocalized (rather than localized as in \cite{R39}). Such a seemingly inconsistency can be resolved by observing that, as shown in \cite{supp}, the non-Hermitian  AAH Hamiltonian in the {\em infinitely-extended} lattice limit can be transformed into the Hatano-Nelson Hamiltonian with incommensurate disorder by a similarity transformation, and that in such a transformation the metal and insulating phases are reversed.\\
The winding number $w$ describes a bulk property of the system, however in our model it is not useful to predict edge states, i.e. to state a bulk-boundary correspondence like in \cite{R39}. Numerical results show that a number of edge states can arise, either at the left or right boundaries, in a lattice with open boundary conditions and in the metallic phase, where $w$ assumes the trivial value $w=0$ (see Fig.S1 in \cite{supp}). The number of edge states sensitively depends on $h$. Since edge states correspond rather generally to eigenstates with complex energies, the $\mathcal{PT}$ symmetric phase is fragile in a system with open boundaries, as already noticed in previous works \cite{R22bis,R22,R23}. Finally, we note that, while the AAH and Nelson-Hatano Hamiltonians are related one another by  a similarity transformation, the non-Hermitian skin effect observed in the latter model, i.e. the shrinking of all bulk states into one edge of the lattice \cite{R37}, is not found in the AAH model. 
\begin{figure}
\includegraphics[width=8cm]{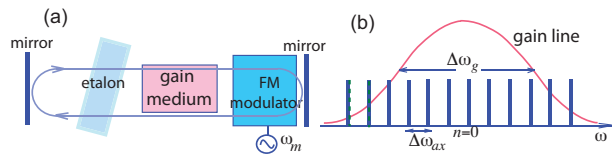}
\caption{(Color online) Non-Hermitian photonic QC. (a) Schematic of the FM mode-locked laser. (b) The axial cavity modes of the laser cavity (vertical lines), coupled by the phase modulator, realize a 1D tight-binding lattice in the spectral domain. The low-finesse etalon introduces an incommensurate complex potential on the lattice.}
\end{figure}

\par

{\it Non-Hermitian photonic quasicrystal.}  
An interesting case is obtained in the double limit $h \rightarrow \infty$, $V \rightarrow 0$ with $V \exp(h) \rightarrow 2 V_0$ finite, corresponding to the incommensurate potential $V_n=V_0 \exp(-2 \pi i \alpha n- i \theta)$ in Eq.(1). In this case, assuming $V_0$ as the control parameter of the Hamiltonian, according to Eq.(7) the topological phase transition is attained as $V_0$ is increased above the critical value $V_{0c}=J$. A simple photonic system that realizes a complex QC of this kind, and thus showing a non-Hermitian topological phase transition, is provided by a frequency-modulated (FM) mode-locked laser \cite{R55,R56}. Mode-locked lasers are routinely used to generate ultrashort optical pulses \cite{R57} and are known to provide experimentally accessible 
systems to observe phase transitions in their spectrum \cite{R58,R59,R60,R61}. A schematic of a mode-locked laser that realizes a non-Hermitian QC is shown in Fig.2(a). It consists of a standard Fabry-Perot laser cavity with axial modes spaced by $\Delta \omega_{ax}$ and with an intracavity phase modulator driven at frequency $\omega_m=\Delta \omega_{ax}$. The gain medium is provided by a homogeneously-broadened active material with a slow relaxation dynamics and wide gain bandwidth $\Delta \omega_g \gg \Delta \omega_{ax}$ \cite{R55}. The spectral axial modes with amplitudes $\psi_n$ realize a 1D lattice in the spectral domain  [Fig.2(b)]. The phase modulator, impressing a time-dependent phase $\Delta \phi(t) = \Delta_{FM} \cos(\omega_m t)$ to the transmitted light field, couples the spectral modes with a hopping rate $J= \Delta_{FM}/2$.  A synthetic complex potential $V_n=V_0 \exp( 2 \pi i \alpha n+ i \theta)$ on the lattice is realized by a low-finesse intracavity etalon, with free spectral range $\Delta \omega_{etal}= \omega_m / \alpha$ incommensurate with respect to the modulation frequency and much smaller than the gain bandwidth. The potential amplitude $V_0$ is determined  by the reflectance $R$ of etalon facets according to the relation $V_0 \simeq R $ (see \cite{supp} for technical details).
The evolution of spectral amplitudes $\psi_n$ at successive round-trips in the cavity is described by the coupled-mode equations of mode-locking in frequency domain \cite{supp,R57,R59}
\begin{equation}
i \frac{d \psi_n}{dt}=J( \psi_{n+1}+\psi_{n-1})+V_0 \exp(2 \pi i \alpha n+ i \theta) \psi_n+ i \mathcal{L} \psi_n
\end{equation}
 where $t$ is the round-trip number (i.e. physical time normalized to the cavity photon transit time) and $\mathcal{L}$ accounts for the effects of cavity losses and amplification in the gain medium. For a homogeneously-broadened gain medium and neglecting other cavity dispersion effects, one can assume \cite{R55,R57}
 $ \mathcal{L} = -\gamma+g/(1+ 4 n^2 \omega_{m}^2 / \Delta \omega_{g}^2)$, where $\gamma$ is the effective loss rate of the cavity per round-trip, $n=0$ is the index of the spectral axial mode at the center of the Lorentzian gainline, and $g$ is the saturated gain. For a slow gain medium, $g$ satisfies the rate equation \cite{R55}
 \begin{equation}
dg/dt= \gamma_{\parallel} \left[ g_0-g(1+I) \right]
 \end{equation} 
 where $g_0$ is the small-signal gain, $\gamma_{\parallel}$ is the relaxation rate of the population inversion normalized to the modulation frequency, and $I= \sum_n |\psi_n|^2$ is the intracavity laser intensity, averaged over the cavity round-trip time and normalized to the saturation intensity of the two-level transition.
 Clearly, in the limit of a broad gainline $\omega_m / \Delta \omega_g  \rightarrow 0$ and assuming $g \simeq \gamma$, the spectral mode dynamics as described by Eq.(8) emulates the non-Hermitian  AAH model. This means that, in the localized (insulating) phase $J<V_0$, a narrow laser spectrum, localized near the center of the gainline, should be observed, whereas a rather abrupt spectral broadening should arise in the delocalized (metallic) phase $J>V_0$. In the latter case  the actual oscillating laser spectrum is ultimately limited by the finite bandwidth of the gain medium according to the Kuizenga-Siegman theory of active mode-locking \cite{R56}. Figure 3 shows a typical behavior of laser spectra, as obtained after transient laser switch-on dynamics starting from random-noise of spectral amplitudes, for increasing values of the FM modulation strength $\Delta_{FM}$. Parameter values used in the simulation are typical for Nd:YAG laser \cite{supp,R56} and correspond to an incommensurate potential with $\alpha=(\sqrt{5}-1)/4$ and $V_0=0.14$. The behavior of the oscillating bandwidth versus $\Delta_{FM}$ [Fig.3(a)] clearly shows an abrupt change at $\Delta_{FM} \simeq 2 V_0$, corresponding to the non-Hermitian topological phase transition from the insulating to the metallic phase.\par
 \begin{figure}
\includegraphics[width=8cm]{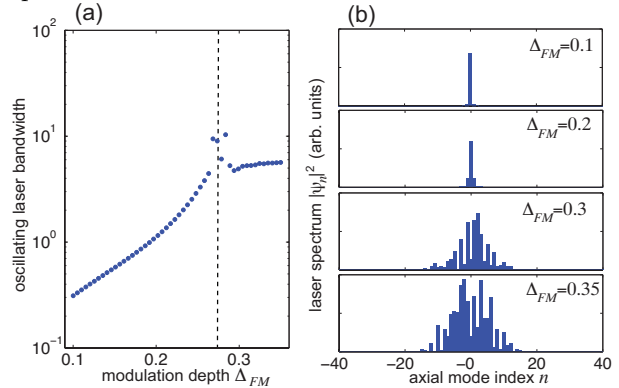}
\caption{Non-Hermitian phase transition in the spectrum of a FM mode-locked laser. (a) Behavior of the spectral width of the oscillating laser modes (in log scale), normalized to the modulation frequency, versus modulation depth $\Delta_{FM}$. (b) Detailed profiles of laser spectra at a few increasing values of $\Delta_{FM}$. The laser spectra are taken after transient relaxation oscillations by numerical simulations of Eqs.(8) and (9). Parameter values used in the simulations are: $\gamma=0.19$, $g_0=3 \gamma$, $\omega_m=2 \pi \times 1.384$ GHz, $\Delta \omega_g= 2 \pi \times  126$ GHz, $\alpha=(\sqrt{5}-1)/2$ and $V_0=0.14$. The dashed vertical curve in (a) corresponds to the critical value $\Delta_{FM}=2 V_0$ of the topological non-Hermitian phase transition.}
\end{figure}

 {\it Conclusions.} 
The discovery of topological phases in non-Hermitian open systems challenges the current wisdom of topological order. Recent studies have provided many insights to solve major issues such as generalizations of the bulk-boundary correspondence, however little is known about topological properties of non-Hermitian QCs. Here we have considered a non-Hermitian extension of the Aubry-Andr\'e-Harper model, and uncovered the topological nature of the non-Hermitian metal-insulator phase transition observed when the complex phase of the incommensurate potential is varied. A photonic realization of a non-Hermitian QC, which could provide a signature of the phase transition, has been also proposed. There are several open questions ahead. For example, the winding number in Eq.(5)  uncovers the bulk topological nature of metallic or insulating phases of the QC, however unlike the Hatano-Nelson crystal with open boundaries, showing the non-Hermitian skin effect \cite{R39}, it is not useful to study edge states. Attempts to provide a topological classification of edge states in non-Hermitian QCs have been suggested very recently \cite{R53palle}. Finally, it would be interesting to extend the present study to other non-Hermitian QCs, such as Fibonacci chains and two-dimensional QCs.
 \par

\vspace{0.5cm}
 \small
$^*$ Corresponding author: stefano.longhi@polimi.it

\begin{center}
\section*{ \bf Supplemental Material}
\end{center}
\renewcommand{\thesubsection}{S}
\renewcommand{\theequation}{S-\arabic{equation}}
\setcounter{equation}{0}

\begin{center}
{\it {\bf S.1. Similarity transformation, $\mathcal{PT}$ symmetry breaking and metal-insulator phase transition}}\par
\end{center}
Let us consider the tight-binding lattice Hamiltonian, given by Eq.(1) in the main text, defined on a ring with $L$ sites and periodic boundary conditions. The integer $L$ is chosen large enough so that
the irrational number $\alpha$ can be approximated, at any arbitrary accuracy, by a rational $\alpha \simeq p/q$ with $p$, $q$ irreducible integers and $q=L$. The thermodynamic limit $L \rightarrow \infty$ should be then taken. The energy spectrum of $H(\varphi)$ is obtained from the eigenvalue equation
\begin{equation}
E \psi_n= J(\psi_{n+1}+\psi_{n-1}) +V \cos (2 \pi \alpha n+ \varphi) \psi_n
\end{equation}
with the periodic boundary conditions
\begin{equation}
\psi_{n+L}= \psi_n
\end{equation}
and involves the calculation of the $L$ eigenvalues and corresponding eigenvectors of the $L \times L$ matrix
\begin{equation}
H(\varphi)=\left(
\begin{array}{cccccccccc}
V_1 & J & 0 & 0 &0 & ... & 0 & 0& J \\
J & V_2 & J & 0 & 0 & ... & 0 & 0 & 0 \\
0 & J  & V_3 & J & 0 & ... & 0 & 0 & 0 \\
... & ... & ... & ... & ... & ... & ... & ...& ...\\
0 & 0 & 0 & 0 & 0 & ... & J & V_{L-1} & J \\
J & 0 & 0 & 0 & 0 & ... & 0 & J & V_L
\end{array}
\right)
\end{equation}
with $V_n=V \cos( 2 \pi \alpha n+ \varphi)$ and $\varphi= \theta+ih$. Let us introduce the discrete Fourier transformation
\begin{equation}
\phi_n= \frac{1}{\sqrt{L}} \sum_{l=1}^L \psi_l \exp(2 \pi i \alpha l n)
\end{equation}
i.e. 
\begin{equation}
\psi_n= \frac{1}{\sqrt{L}} \sum_{l=1}^L \phi_l \exp(-2 \pi i \alpha l n).
\end{equation}
The eigenvalue equation (S-1) is then transformed into the following one
\begin{equation}
E \phi_n=\frac{V}{2} \{ \exp(i \varphi) \phi_{n+1}+\exp(-i \varphi) \phi_{n-1} \} +W_n \phi_n
\end{equation}
where we have set
\begin{equation}
W_n= 2 J \cos (2 \pi \alpha n)
\end{equation}
and where the periodic boundary conditions $\phi_{n+L}=\phi_n$ hold. 
Let us focus our attention to the case where $H$ is $\mathcal{PT}$ symmetric, i.e. let us assume $\theta=0$. Interestingly, Eq.(S-6) describes the 
Hatano-Nelson model [1-4] with incommensurate disorder $W_n$ of on-site energies on a ring, the imaginary gauge field $h$ being determined by the complex phase $\varphi=ih$. Clearly, the eigenvalue problem (S-6)  involves the determination of the eigenenergies and eigenstates of the $L \times L$ matrix
\begin{widetext}
\begin{equation}
H_1=\left(
\begin{array}{cccccccccc}
W_1 & \frac{V}{2} \exp(-h) & 0 & 0 &0 & ... & 0 & 0& \frac{V}{2} \exp(h) \\
\frac{V}{2} \exp(h) & W_2 & \frac{V}{2} \exp(-h) & 0 & 0 & ... & 0 & 0 & 0 \\
0 & \frac{V}{2} \exp(h)  & W_3 & \frac{V}{2} \exp(-h) & 0 & ... & 0 & 0 & 0 \\
... & ... & ... & ... & ... & ... & ... & ...& ...\\
0 & 0 & 0 & 0 & 0 & ... & \frac{V}{2} \exp(h)  & W_{L-1} & \frac{V}{2} \exp(-h) \\
\frac{V}{2} \exp(-h)  & 0 & 0 & 0 & 0 & ... & 0 & \frac{V}{2} \exp(h)  & W_L
\end{array}
\right)
\end{equation}
\end{widetext}
Note that $H$ and $H_1$, defined by Eqs.(S-3) and (S-8), can be obtained one another by the similarity transformation $H_1= T H T^{-1}$ where $T_{n,l}=(1/ \sqrt{L}) \exp( 2 \pi i \alpha l n)$. For $h=0$, $H$ and $H_1$ are equivalent one another after the change $V \leftrightarrow 2J$, which is the well-known duality property of the Hermitian AAH model (self-duality is attained as $V=2J$ [5,6]). In the non-Hermitian case $h \neq 0$, the duality property is violated, however the two matrices $H$ and $H_1$  are similar and thus share the same set of eigenvalues. Moreover, in the $L \rightarrow \infty$ limit, a localized eigenstate of $H$ corresponds to a delocalized eigenstate of $H_1$, and viceversa. This property readily follows from Eqs.(S-4) and (S-5), that relate $\psi_n$ and $\phi_n$ by a discrete Fourier transform. Therefore, the energy spectrum and localization properties of the eigenstates of the Hamiltonian $H$ can be retrieved from those of the Hamiltonian $H_1$. Let us assume $V<2J$, so that for $h=0$ all eigenstates $\phi_n$ of $H_1$  are localized (and thus those of $H$ delocalized), with an inverse localization length (Lyapunov exponent)  $\gamma$ which is independent of the eigenenergy and given by $\gamma=\log(2J/V)$ {[5,6]}. According to the Hatano-Nelson conjecture [1-4], as $h$ is adiabatically increased from zero, in the thermodynamic limit $L \rightarrow 0$ the eigenenergies of $H_1$ are not changed, all eigenstates $\phi_n$ remain localized and are obtained from those of the Hermitian problem $h=0$ after multiplication by the exponential factor $\sim \exp(hn)$. As  $h$ is increased to reach and overcome the critical value $h_c=\gamma=\log(2J/V)$, the exponential tail term $\exp(hn)$ dominates over the exponential decay tail $\exp(-\gamma |n|)$ of $\phi_n$ in the Hermitian limit, so that a phase transition arises: at $h>h_c$ all eigenstates of $H_1$ become delocalized and the energy spectrum complex. This phase transition is analogous to the non-Hermitian delocalization transition originally predicted by Hatano and Nelson for the problem of Anderson localization [1-3].  Note that, while in the Anderson problem (random on-site potential disorder)  mobility edges separating localized and delocalized states are observed for any non-vanishing value of $h$, for incommensurate on-site disorder there are not mobility edges. The reason thereof is that in the Anderson model there are weakly localized eigenstates near the band center, where even a small non vanishing imaginary gauge field opens a narrow mobility region, while in the incommensurate potential problem all eigenstates have the same localization length and thus they are {\em all} localized ($h<h_c$) or delocalized ($h>h_c$).\\
For the Hamiltonian $H$, the non-Hermitian delocalization transition observed for $H_1$ means that: (i) for $h<h_c$ the energy spectrum of $H$ remains real and does not depend on $h$, i.e. $H$ is in the unbroken $\mathcal{PT}$ phase. Also, all eigenstates of $H$ are delocalized (metallic phase); (ii) for $h>h_c$, the energy spectrum of $H$ becomes complex (unbroken $\mathcal{PT}$ phase) and all eigenstates of $H$ become localized (insulating phase). This proves that the $\mathcal{PT}$ symmetry breaking phase transition and the metal-insulating phase  
transition coincide for the non-Hermitian AAH model and occur at the critical value $h=h_c$ of the complex phase given by Eq.(7) in the main text.\\
\\
\begin{center}
{\it {\bf S.2. Winding number}}\par
\end{center}
The $\mathcal{PT}$ symmetry breaking and metal-insulator phase transitions occurring at $h=h_c$ have a topological nature, and can be traced back to an abrupt change of a winging number $w=w(h)$ associated to  $H(\varphi)$ as $h$ is varied from below to above the critical value $h_c$. To properly define the winding number $w$, let us assume $V< 2J$, so that for $h=0$ the spectrum of $H(\varphi)$ is absolutely continuous, independent of $\theta$, and has a Cantor-set structure with dense gaps [7,8]. The measure of the bands approaches zero as $V$ approaches its dual value $2J$ from below. Let us indicate by $E_B$ a base energy, chosen to be in a small gap of the Cantor-set. For $h>0$, let us introduce the winding number $w$ as follows
\begin{eqnarray}
w(h) & = & \lim_{L \rightarrow \infty }\frac{1}{2 \pi i} \int_0^{2 \pi} d \theta \frac{\partial}{\partial \theta} \log  \left\{ \det \left(  H\left( \frac{\theta}{L},h \right)-E_B \right) \right\} \nonumber \\
& = & \lim_{L \rightarrow \infty} \int_0^{2 \pi}  d \theta  \frac{1}{2 \pi i f}\frac{\partial f}{\partial \theta} 
\end{eqnarray}
where we have set
\begin{equation}
f(\theta,h) \equiv  \det \left\{ H\left( \frac{\theta}{L},h \right)-E_B \right\}.
\end{equation}
The main result that we are going to demonstrate is that $w$ is independent of $E_B$ and takes the value $w=0$ for $h<h_c$, and $w=-1$ for $h>h_c$, where $h_c= \log (2J/V)$ is the critical value of the imaginary phase corresponding to the metal-insulator ($\mathcal{PT}$ symmetry breaking) phase transition. To calculate $f$, let us notice that the matrix $H(\theta/L,h)$ can be transformed, via a similarity transformation $H=R^{-1} H_2 R$ with $R_{n,l}=(1/\sqrt{L}) \exp( 2 \pi i \alpha n l) \exp(-nh+i n \theta/L)$, into the following matrix
\begin{widetext}
\begin{equation}
H_2=\left(
\begin{array}{cccccccccc}
W_1 & \frac{V}{2} & 0 & 0 &0 & ... & 0 & 0& \frac{V}{2} \exp(hL-i \theta) \\
\frac{V}{2} & W_2 & \frac{V}{2}  & 0 & 0 & ... & 0 & 0 & 0 \\
0 & \frac{V}{2}  & W_3 & \frac{V}{2} & 0 & ... & 0 & 0 & 0 \\
... & ... & ... & ... & ... & ... & ... & ...& ...\\
0 & 0 & 0 & 0 & 0 & ... & \frac{V}{2}  & W_{L-1} & \frac{V}{2}\\
\frac{V}{2} \exp(i \theta-hL)  & 0 & 0 & 0 & 0 & ... & 0 & \frac{V}{2} & W_L
\end{array}
\right)
\end{equation}
\end{widetext}
with $W_n= 2 J \cos (2 \pi \alpha n)$. Hence
\begin{equation}
f(\theta,h) =  \det ( H_2-E_B).
\end{equation}
To calculate the determinant of $(H_2-E_B)$, let us notice that, in the large $L$ limit and for $h>0$, the element $(H_2)_{L,1}= (V/2) \exp(i \theta-hL)$ exponentially vanishes and can be neglected. This yields
\begin{equation}
f(\theta,h)=(-1)^{L+1}\left( \frac{V}{2} \right)^{L} \exp(hL-i \theta)+ \det (\Theta-E_B)
\end{equation}
where the  Hermitian matrix $\Theta$ is defined by
\begin{widetext}
\begin{equation}
\Theta=\left(
\begin{array}{cccccccccc}
W_1& \frac{V}{2} & 0 & 0 &0 & ... & 0 & 0& 0 \\
\frac{V}{2} & W_2 & \frac{V}{2}  & 0 & 0 & ... & 0 & 0 & 0 \\
0 & \frac{V}{2}  & W_3 & \frac{V}{2} & 0 & ... & 0 & 0 & 0 \\
... & ... & ... & ... & ... & ... & ... & ...& ...\\
0 & 0 & 0 & 0 & 0 & ... & \frac{V}{2}  & W_{L-1} & \frac{V}{2}\\
0  & 0 & 0 & 0 & 0 & ... & 0 & \frac{V}{2} & W_L
\end{array}
\right)
\end{equation}
\end{widetext}
i.e. it is obtained from $H_2$ after letting $(H_2)_{1,L}=(H_2)_{L,1}=0$
Clearly, the matrix $\Theta$ describes the Hermitian AAH model  in the insulating phase ($V<2J$) on a linear chain with $L$ sites and open boundary conditions (this is because the corner elements $\Theta_{1L}$ and $\Theta_{L1}$ of $\Theta$ vanish). Since $\Theta$ does not depend on $\theta$, from Eq.(S-13) one obtains
\begin{equation}
\frac{1}{2 \pi i f} \frac{\partial f}{\partial \theta}=\frac{1}{2 \pi} \frac{(-1)^{L}\left( \frac{V}{2} \right)^{L} \exp(hL-i \theta)}{(-1)^{L+1}\left( \frac{V}{2} \right)^{L} \exp(hL-i \theta)+ \det (\Theta-E_B)}
\end{equation}
To calculate the limit of Eq.(S-15) as $L \rightarrow \infty$, we need to estimate the behavior of $| \det(\Theta-E_B)|$ as $L \rightarrow \infty$. Indicating by $\lambda_1$, $\lambda_2$, ..., $\lambda_L$ the $L$ eigenvalues of $\Theta$ and setting $g \equiv \log |\det (\Theta-E_B)|$, one has
\begin{equation}
g=\sum_{l=1}^L \log |\lambda_l-E_B|.
\end{equation}
Note that, since $E_B$ is chosen in a small gap of the absolutely continuous energy spectrum, $E_B \neq \lambda_l$ and $g$ does not diverge.
In the large $L$ limit Eq.(S-16) can be written as
\begin{equation}
g=L \int d \epsilon \rho( \epsilon) \log |\epsilon-E_B|
\end{equation}
where $\rho(\epsilon)$ is the density of states of $\Theta$ at energy $\epsilon$ defined in the usual way, i.e. $L \rho( \epsilon)d \epsilon$ is the number of eigenvalues $\lambda_l$ of $\Theta$ is the interval $(\epsilon, \epsilon+ d \epsilon)$. According to Thouless [9], the inverse of the localization length of an eigenstate of $\Theta$ with an energy in the neighborhood of $E_B$ is given by
\begin{equation}
\gamma= \int d \epsilon \rho( \epsilon) \log | \epsilon-E_B|- \log \left( \frac{V}{2} \right).
\end{equation}
Therefore, from Eqs.(S-17) and (S-18) one obtains
\begin{equation}
|\det ( \Theta-E_B) |=\exp (g)=\left( \frac{V}{2} \right)^L \exp( \gamma L).
\end{equation}
On the other hand, since $\Theta$ is the matrix Hamiltonian of the Hermitian AAH model in the insulating (localized) phase, for the duality property it is well known that the inverse of the localization length $\gamma$ does not depend on the energy $E_B$ and is equal to 
\begin{equation}
\gamma=\log (2J/V)
\end{equation}
 for all eigenstates [5,6]. From Eqs.(S-15) and (S-19), it then readily follows that
\begin{equation}
\lim_{L \rightarrow \infty} \frac{1}{2 \pi i f} \frac{\partial f}{\partial \theta}= \left\{ 
\begin{array}{cc}
0 & h<\gamma \\
-(1/ 2 \pi)  & h> \gamma
\end{array}
\right.
\end{equation}
Finally, substitution of Eq.(S-21) into Eq.(S-9) yields the main result $w(h)=0$ for $h< \gamma=h_c$ and $w=-1$ for $h>h_c$.\par
\begin{center}
\bf{S.3 Edge effects}
\end{center}
\begin{figure*}
\includegraphics[width=17cm]{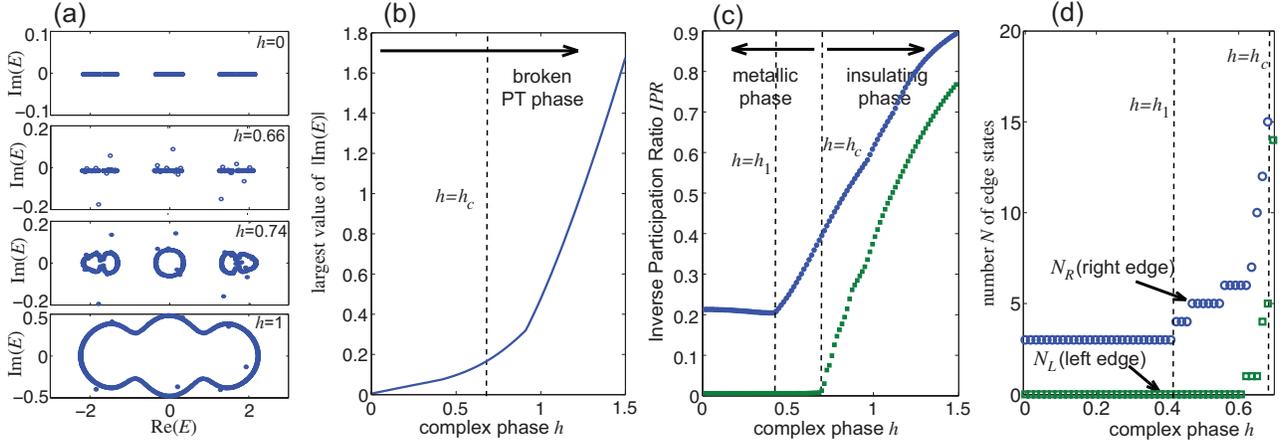}
\caption{(Color online) Edge effects in a non-Hermitian AAH  Hamiltonian $H(\theta,h)$ with open boundary conditions for $J=V=1$, $\alpha=(\sqrt{5}-1)/2$ and $L=500$. (a) Numerically-computed energy spectrum $E$ of $H$ for $\theta=0$ and for a few increasing values of the complex phase $h$. (b) Behavior of the largest value of $|{\rm Im}(E)|$ versus the non-Hermitian phase $h$ for $\theta=0$. The dashed vertical curve corresponds to the critical value $h_c={\rm log}(2J/V) \simeq 0.6931$ of metal-insulator phase transition. (c) Numerically-computed behavior of the largest and smallest values of inverse participation ratio IPR of eigenstates versus $h$. (d) Numerically-computed number of edge states, in the left ($N_L$) and right ($N_R$) edges of the chain, versus $h$ in the metallic phase $h<h_c$. For $h<h_1 \simeq 0.41$ there are 3 edge states on the right edge and zero edge states on the left edge.}
\end{figure*}
Let us consider the non-Hermitian AAH model on a finite lattice with {\it open} boundary conditions. The Hamiltonian of the system is now described by Eq.(S-3) but with the corner elements $H_{1,L}$ and $H_{L,1}$ of $H(\varphi)$ replaced by zero. Owing to lattice truncation, edge states can arise, either on the left or right edges of the lattice, corresponding rather generally to complex energy eigenstates when $h \neq 0$ [10-12]. This makes the $\mathcal{PT}$ symmetric phase fragile in lattices with open boundary conditions, i.e. edge effects bring the system into the broken $\mathcal{PT}$ phase for a nonvanishing complex phase $h$. This is clearly illustrated in Fig.4(a) and (b), where the energy spectrum of $H$ for increasing values of $h$ is shown in the incommensurate potential with open boundary conditions. Note that the eigenenergy with the largest imaginary part (in modulus) monotonously increases with $h$, indicating that the system rapidly enters into the broken $\mathcal{PT}$ phase.  The behavior of the IPR of eigenstates (largest and smallest value of IPR) versus $h$ is shown in Fig.4(c). Like in Fig.1(c) of the main text, for $h>h_c$ all states become localized, i.e. a metal-insulator phase transition is observed. However, with open boundary conditions edge states are found in the metallic phase $h<h_c$. For the chosen parameter values, at $h=0$ the Hamiltonian sustains three edge states on the right edge, and zero edge states on the left edge. Figure 4(d) shows the behavior of the number of edge states $N_L$ and $N_R$ on the left and right boundaries for increasing values of $h$ below the critical value $h_c$. As $h$ is increased from zero, below the value $h_1 \simeq 0.41 < h_c$ the  number of edge states remains unchanged, whereas for $h>h_1$ the number of edge states increases and also appear on the left edge, until the system enters into the insulating phase and all eigenstates become localized.\\
\\
\begin{center}
{\it {\bf S.4. Non-Hermitian quasicrystal in a FM mode-locked laser}}\par
\end{center}  
A simple photonic system that implements a QC  with an incommensurate complex potential of the form
\begin{equation}
V_n =V_0 \exp( 2 \pi i \alpha n+ i \theta)
\end{equation}
is provided by a FM mode-locked laser schematically shown in Fig.2(a) of the main text. As the light field circulates back and forth between the two end mirrors of the cavity, the spectral components $\psi_n$ of cavity axial modes undergo small changes when crossing the gain medium, the phase modulator, and the etalon. Indicating by $t$ the round-trip number in the cavity, i.e. the physical time normalized to the photon transit time in the resonator, one can write [13]
\begin{eqnarray}
\psi_n(t+1) & = & \psi_n(t)+ \{ \delta \psi_n(t)\}_{gain}+\{ \delta \psi_n(t)\}_{loss} \;\;\;\;\;\;\;  \nonumber \\
& + & \{ \delta \psi_n(t)\}_{modul}+\{ \delta \psi_n(t)\}_{etalon} 
\end{eqnarray} 
where $\{ \delta \psi_n(t)\}_{gain}$ is the small change of $\psi_n$ due to the round-trip passage in the gain medium (spectral mode amplification), $\{ \delta \psi_n(t)\}_{loss}=-l \psi_n$ accounts for the cavity losses with loss rate $l$ per round-trip, $\{ \delta \psi_n(t)\}_{modul}$ is the change introduced by the phase modulator, and $\{ \delta \psi_n(t)\}_{etalon}$ is the change arising from etalon transmission. For a homogeneously-broadened gain medium with Lorentzian lineshape and neglecting dispersion effects of the gain medium, one can  write [13]
\begin{equation}
\{ \delta \psi_n(t)\}_{gain}=\frac{g}{1+4 n^2 \omega_m^2 / \Delta \omega_g^2}
\end{equation}
where $g$ is the saturated gain and $\Delta \omega_g$ the full-width at half maximum (FWHM) of the gainline. The phase modulator introduces a time-dependent phase shift $\Delta_{FM} \cos( \omega_m t)$ to the incident light field, so that for a small modulation depth $\Delta_{FM} \ll 1$ in the spectral domain one has
\begin{equation}
\{ \delta \psi_n(t)\}_{modul}=-i \frac{\Delta_{FM}}{2} (\psi_{n+1}+\psi_{n-1}).
\end{equation}
To calculate $\{ \delta \psi_n(t)\}_{etalon}$, let us assume that the etalon is uncoated and made of a transparent glass of refractive index $n_1$ and thickness  $L$, placed inside the cavity and slightly tilted from normal incidence. Indicating by $t_{et}(\omega)$ the spectral transmission amplitude of the etalon, with $\omega=0$ taken at the center of the gainline, one has
\begin{equation}
\{ \delta \psi_n(t)\}_{etalon} =[t_{et}(n \omega_{m})-1] \psi_n(t)
\end{equation}  
For near-normal incidence, the spectral transmission of a non-absorbing etalon reads [14]
\begin{equation}
t_{et}(\omega)=\frac{1-R}{1-R \exp(2 i \delta+i \phi)}
\end{equation}
where $\delta=n_1 L \omega / c$, $c$ is the speed of light in vacuum, $\phi$ is an additional phase shift that can be adjusted by slight etalon tilting, and
\begin{equation}
R \simeq \left( \frac{n_1-1}{n_1+1} \right)^2
\end{equation}
 is the Fresnel reflectance of the air/glass interface at near-normal incidence. Since $R \ll 1$ for a glass-air interface, at first-order approximation one can write
\begin{equation}
t_{et}(\omega) \simeq 1-R+ R \exp(2 i \delta+i \phi)
\end{equation}
so that one obtains
\begin{equation}
\{ \delta \psi_n(t)\}_{etalon} =[-R+ R \exp(2 \pi i \alpha n+ i \phi)] \psi_n(t)
\end{equation}  
where $\alpha= \omega_m / \Delta \omega_{etal}$ and $\Delta \omega_{etal}=\pi c /(L n_1)$ is the free-spectral-range of the etalon. Substitution of Eqs.(S-24,S-25,S-30) into Eq.(S-23) and after setting $\psi_n(t+1)-\psi_n(t) \simeq ( \partial \psi_n/ \partial t)$, one obtains
\begin{equation}
i \frac{\partial \psi_n}{\partial t}=J \left( \psi_{n+1} +\psi_{n-1} \right) + i \mathcal{L} \psi+V_0 \exp( 2 \pi i \alpha n +i \theta) \psi_n
\end{equation}
where we have set $J = \Delta_{FM}/2$, $\mathcal{L} \equiv -\gamma+g/ (1+ 4 n^2 \omega_m^2 / \Delta \omega_g^2)$, $\gamma=l+R$, $V_0=R$ and $\theta= \phi+\pi/2$. Equation (S-31) is precisely Eq.(8) given in the main text.\\
\begin{figure}
\includegraphics[width=8cm]{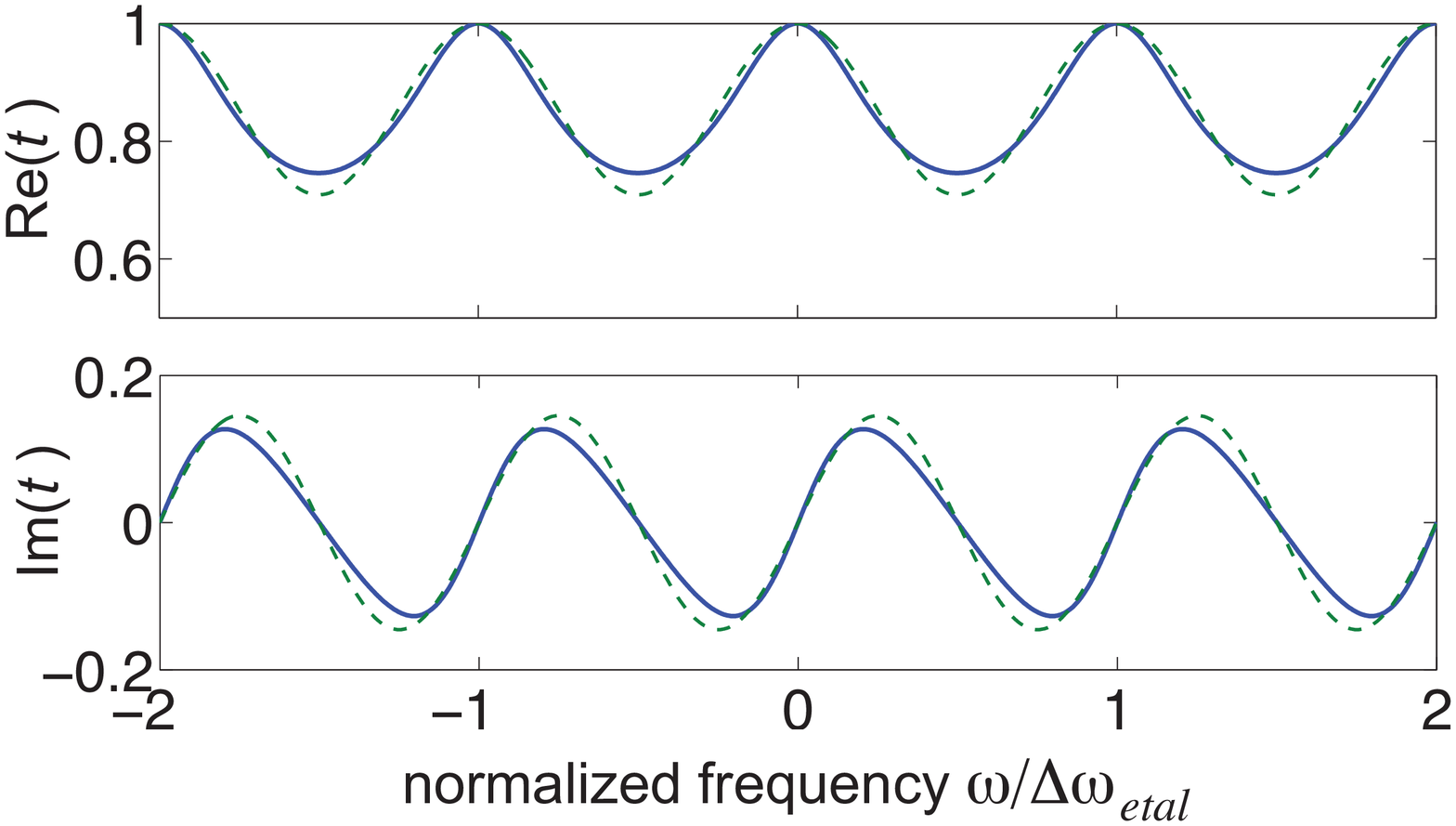}
\caption{(Color online) Spectral transmission (real and imaginary parts of the transmission amplitude $t$) in an uncoated etalon at near normal incidence for $L=3$ cm and $n_1=2.2321$. Solid and dashed curves refer to the exact and approximate relations given by Eq.(S-27) and (S-29), respectively.}
\end{figure}
In an experimental setting, there is no need to place an etalon in the laser cavity, since etalon effects can arise from the low Fresnel reflectivity of the two plane facets of the nonlinear crystal used in the FM modulator, as demonstrated in [15].  To provide physical parameters, let us consider a typical FM mode-locked Nd:YAG laser [15] emitting at $\lambda_0=1064$ nm wavelength. Assuming a LiNbO$_3$-based phase modulator with crystal thickness $L=3 $ cm and refractive index $n_1=2.2321$ at laser wavelength, the free spectral range of the etalon realized by the two plane facets of the nonlinear crystal is $\Delta \omega_{etal}= 2 \pi \times 2.24$ GHz and the amplitude of the potential is $V_0=R \simeq 0.1453$ according to Eq.(S-28). The behavior of the spectral etalon transmission $t_{et}$ versus frequency, as obtained from the exact relation (S-27) and first-order approximation (S-29), is shown in Fig.5. The irrational value $ \omega_m / \Delta \omega_{et}=(\sqrt{5}-1)/2$ is obtained by driving the phase modulator at the frequency $\omega_m \simeq 2 \pi \times 1.3844$ GHz, corresponding to an optical cavity length of the laser system of $ \simeq 10.84$ cm. The linewidth of the homogeneously-broadened gainline is $\Delta \omega_g=2 \pi \times 126$ GHz, whereas the relaxation rate of population inversion is $\sim 1/230$ MHz. Such physical parameters have been used in numerical simulations of the laser equations (Fig.3 in the main text).

 \vspace{20px}
\noindent
\small
{[1]} N. Hatano and D. R. Nelson, Localization Transitions in Non-Hermitian Quantum Mechanics, Phys. Rev. Lett. {\bf 77}, 570 (1996).\\
{[2]} N. Hatano and D. R. Nelson, Vortex pinning and non-Hermitian quantum mechanics, Phys. Rev. B {\bf 56}, 8651 (1998).\\
{[3]} N. Hatano and D. R. Nelson, Non-Hermitian Delocalization and Eigenfunctions, Phys. Rev. B {\bf 58}, 8384 (1998).\\
{[4]} P. W. Brouwer, P. G. Silvestrov, and C.W.J. Beenakker, Theory of Directed Localization in one dimension, Phys. Rev. B {\bf 56}, 4333-4335 (1997).\\
{[5]} S. Aubry and G. Andr\'e, Analyticity breaking and Anderson localization in incommensurate lattices, Ann. Israel Phys. Soc. {\bf 3}, 133-140 (1980).\\
{[6]} J.B. Sokoloff, Unusual band structure, wave function and electrical conductance in crystals with incommensurate periodic potentials, Phys. Rep. {\bf 126}, 189-244 (1984).\\
{[7]} S. Ostlund and R. Pandit, Renormalization-group analysis of the discrete quasiperiodic Schrodinger equation, Phys. Rev. B {\bf 29}, 1394 (1984).\\
{[8]} A. Avila and S. Jitomirskaya, The Ten Martini Problem,  Ann. Math. {\bf 170} , 303 (2009).\\
{[9]}  D.J. Thouless, A relation between the density of states and range of localization for one dimensional random systems, J. Phys. C: Solid State Phys. {\bf 5}, 77 (1973).\\
{[10]} S. Longhi, $\mathcal{PT}$-symmetric optical superlattices, J. Phys. A {\bf 47}, 165302 (2014).\\
{[11]}C. Yuce, $\mathcal{PT}$-symmetric Aubry-Andr\'e model, Phys. Lett. A {\bf 378}, 2024 (2014).\\
{[12]} C.H. Liang, D.D. Scott, and Y.N. Joglekar, $\mathcal{PT}$ restoration via increased loss-gain in $\mathcal{PT}$ -symmetric Aubry-Andr\'e model, Phys. Rev. A {\bf 89}, 030102 (2014).\\
{[13]} H.A. Haus, Mode-locking of lasers, IEEE J. Sel. Top. Quantum Electron. {\bf 6}, 1173 (2000).\\
{[14]} D.J. Kuizenga and A.E. Siegman, FM and AM Mode Locking of the Homogeneous Laser-Part I, IEEE J. Quantum Electron. {\bf 6}, 694 (1970).\\
{[15]} D.J. Kuizenga and A.E. Siegman, FM and AM Mode Locking of the Homogeneous Laser-Part II, IEEE J. Quantum Electron. 6, 709 (1970).\\

\end{document}